\documentclass[superscriptaddress,letterpaper,preprint]{revtex4}
\usepackage{amsmath}

\def\beq{\begin{equation}}
\def\eeq{\end{equation}}
\def\beqa{\begin{eqnarray}}
\def\eeqa{\end{eqnarray}}
\def\ban{\begin{eqnarray*}}
\def\ean{\end{eqnarray*}}
\def\bi{\begin{itemize}}
\def\ei{\end{itemize}}

\begin{document}

\title{Center-of-Mass Correction in a Relativistic Hartree
Approximation Including the Meson Degrees of Freedom}

\author{P. Alberto}
\affiliation{Center for Computational Physics and Physics Department, University of Coimbra\\
P-3004-516 Coimbra, Portugal}

\author{S.S. Avancini}
\affiliation{Depto de F\'{\i}sica, CFM, Universidade Federal de Santa
Catarina, Florian\'opolis,  SC,  CP. 476,  CEP 88.040 - 900,  Brazil}

\author{M. Fiolhais}
\affiliation{Center for Computational Physics and Physics Department, University of Coimbra\\
P-3004-516 Coimbra, Portugal}

\author{J.R. Marinelli}
\affiliation{Depto de F\'{\i}sica, CFM,  Universidade Federal de Santa
Catarina,  Florian\'opolis,  SC,  CP. 476,  CEP 88.040 - 900,  Brazil}

\begin{abstract}
We use the Peierls-Yoccoz projection method to study the motion of
a relativistic system of nucleons interacting with sigma and omega
mesons, generalizing a method developed for the alpha particle.
The nuclear system is described in a mean-field Hartree approach,
including explicitly the meson contribution. The formalism is
applied to $^{4}{\rm He}$, $^{16}{\rm O}$ and $^{40}{\rm Ca}$. The
center-of-mass correction makes the system too much bounded. It
turns out that a new set of model parameters is needed when the
center-of-mass motion is consistently treated with respect to the
traditional approaches. An appropriate refitting of the model
brings the radii and binding energies to reasonable values for the
oxygen and calcium.
\end{abstract}

\maketitle \vspace{0.30cm} PACS number(s): 21.10.-k, 21.10.Dr,
21.60.-n

\vspace{0.30cm} \baselineskip1.1cm

\section{Introduction}

Relativistic models for finite nuclei, with nucleons and mesons are
usually treated in the Hartree or Hartree-Fock approximations. In
these approximations the total linear momentum is no longer a
conserved quantity and the spurious center of mass (CM) motion gives
rise to unphysical contributions in the calculated nuclear
observables. As already discussed for non-relativistic nuclear
models, using the same kind of mean-field approximation, the correct
treatment for the CM motion introduces a modest modification in the
total binding energy for intermediate mass nuclei, but relatively
large contributions in other observables, e.g., charge distributions
and spectral functions \cite{Schmid1},\cite{Schmid2}. In
relativistic treatments, the CM correction is up to now limited to
the harmonic approximation for the energy or to the subtraction of
$\frac{<\hat{\vec{P^{2}_{A}}}>}{2AM}$ from the total energy, where
$\hat{\vec{P_{A}}}$ is the total nucleonic momentum operator, $M$ is the
nucleon mass  and the mean-value is taken using the Hartree
self-consistent state for $A$ nucleons.

More recently \cite{BJP}, the CM energy correction was estimated
within the $\sigma-\omega$ model, using the Peierls-Yoccoz
projection procedure \cite{Greiner}, for $N=Z$ spherical nuclei
within the Hartree approximation. Although the correction obtained
in this way is of the same order of magnitude of the harmonic
approximation, only the nucleonic degrees of freedom were taken into
account in that calculation. In reference \cite{Alberto}, a
formalism has been developed to include the mesonic degrees of
freedom in the CM projection within $\sigma-\omega$ models and an
application was then made to the $^{4}$He nucleus. In the present
paper we generalize the results obtained in \cite{Alberto} in order
to extend the calculations to heavier spherical nuclei, allowing us
to draw more systematic conclusions. In section II, we review the
main results from Ref. \cite{Alberto}. Then, in section III, the
linear momentum projection within the model is presented and the
total energy functional is worked out. Since most of the model
parametrizations within $\sigma-\omega$ models are based on fits to
the experimental data of both binding energy and charge radius, we
take the same point of view. The nuclear charge mean square radius
is discussed in section IV. The numerical results for $^{4}{\rm
He}$, $^{16}{\rm O}$ and $^{40}{\rm Ca}$ are shown and discussed in
section V. Finally the conclusions and perspectives are summarized
in section VI.

\vspace{0.30cm} \baselineskip1.1cm

\section{The mean-field Hamiltonian with Mesons \\ as Coherent States}
In this section we summarize the main aspects of the relativistic
nucleon-meson models of nuclei. The model used in this work is
restricted to  $\sigma$ and $\omega$ mesons, without
self-interactions (the inclusion of such interactions is
straightforward using the method presented in this paper).

The Lagrangian density for a system of nucleons interacting with
sigma and omega mesons reads \cite{Serot}
\begin{eqnarray}
L &=&L_{\rm N}^{\rm free}+L_\sigma ^{\rm free}+L_\omega ^{\rm
free}+L_{{\rm N}{\rm N}\sigma }^{\rm int}+L_{{\rm N}{\rm N}\omega
}^{\rm int},
\end{eqnarray}
where  N denotes the nucleon and $\sigma $, $\omega $ the mesons.
The Lagrangians for the free fields are:
\begin{eqnarray}
L_{\rm N}^{\rm free} &=&\overline{\psi }(x)({\rm i} \gamma ^\mu \partial _\mu +M)\psi (x), \\
L_\sigma ^{\rm free} &=&-\frac 12\left[ m_\sigma ^2\sigma
^2(x)-\partial _\mu
\sigma (x)\partial ^\mu \sigma (x)\right], \\
L_\omega ^{\rm free} &=&-\frac 14F_{\mu \nu }(x)F^{\mu \nu
}(x)+\frac 12 m_\omega ^2\, \omega _\nu (x)\, \omega ^\nu (x),
\end{eqnarray}
where
\[
F_{\mu \nu }\equiv \partial _\mu \omega _\nu (x)-\partial _\nu
\omega _\mu (x)\, ,
\]
$M$ is the rest mass of the nucleon, and $m_\sigma$ and $m_\omega $
are the meson masses. The sigma
 and omega fields are denoted respectively by $\sigma (x)$ and
$\omega ^\nu (x)$, and the nucleon field by $\psi (x)$. The
interaction parts of the Lagrangian are
\begin{eqnarray}
L_{{\rm N}{\rm N}\sigma }^{\rm int} &=&g_\sigma \overline{\psi }(x)\sigma (x)\psi (x) \\
L_{{\rm N}{\rm N}\omega }^{\rm int} &=&-g_\omega \overline{\psi
}(x)\omega ^\nu (x)\gamma _\nu \psi (x)\, .
\end{eqnarray}

From the above Lagrangian density one derives the following
Hamiltonian density:
\begin{eqnarray}
\mathcal{H}=\mathcal{H}_{N}+\mathcal{H}_{\omega} +
\mathcal{H}_{\sigma}
\label{entot}
\end{eqnarray}
where the fermionic term is
\begin{eqnarray}
\mathcal{H}_{\rm N}&=& \psi^{\dagger}(x)\left\{ -{\rm i}
{\vec{\alpha}}\cdot\vec{\nabla}+\beta\left[M-g_\sigma\sigma(x)\right]-g_\omega\vec{\alpha}\cdot\vec{\omega}(x)+
\frac{g_\omega}{m_{\omega}^{2}}\vec{\nabla} \cdot
\vec{P_{\omega}}\right\} \psi(x)\nonumber \\
& &
+\frac{g_\omega^{2}}{2m_{\omega}^{2}}\left[\psi^{\dagger}(x)\psi(x)\right]^{2}
, \label{toth}
\end{eqnarray}
\noindent and the free meson terms are:
\begin{eqnarray}
\mathcal{H}_{\omega}=\frac{1}{2}\left[\vec{P_{\omega}}\cdot\vec{P_{\omega}}+\frac{(\vec{\nabla}\cdot\vec{P_{\omega}})^{2}}
{m_{\omega}^{2}}+(\vec{\nabla}\times\vec{\omega})^{2}+m_{\omega}^{2}\vec{\omega}^{2}\right]~,\label{sigh}
\end{eqnarray}
\begin{eqnarray}
\mathcal{H}_{\sigma}=\frac{1}{2}[P_{\sigma}^{2}+\vec{\nabla}\sigma\cdot\vec{\nabla}\sigma+m_{\sigma}^{2}\sigma^{2}]~.\label{omegh}
\end{eqnarray}
 Note that we have used the definitions
$P_{\omega}^{i}=F^{0i}$ with $i=1,2,3$ and
$P_{\sigma}=\partial_{0}\sigma$. The quantization of the model
follows the usual procedure described, for instance,  in
\cite{TDLee}. The most important steps of the quantization are
described in \cite{Alberto}, but the procedure can be be outlined
here by saying that, for the canonical quantization of massive
vector fields, one cannot use the field $\omega_0$, because its
canonical conjugate field is identically zero. For conserved
four-vector sources (as it is the case of the nucleon vector
current) the four-divergence of $\omega_\mu$ is zero, and therefore
one can use the full Klein-Gordon equation for $\omega_0$ to write
this field in terms of the divergence of the conjugate field of the
spatial components $\omega^i$ and of the zero-th component of the
vector current, $g_\omega\psi^\dagger\psi$ (see eq. (12) in Ref.
\cite{Alberto}). The Hamiltonian is then built in the usual way by
using only the  spatial components, $\omega^i$, and their respective
conjugate fields, $P_\omega^i$. The two-body contact term in
(\ref{toth}) arises from the quadratic term in $\omega_0$ in the
Lagrangian.

 The nucleon field operators can be expanded as
\begin{eqnarray}
\hat{\psi} (x) &=&\displaystyle\sum_\alpha u_\alpha (\vec{r})\, {\rm
e}^{-{\rm i}  E_\alpha t}\, b_\alpha +\displaystyle\sum_\alpha
v_\alpha (\vec{r})\, {\rm e}^{{\rm i}E_\alpha
t}\, d_\alpha ^{\dagger } \\
\hat{\psi} ^{\dagger }(x) &=&\displaystyle\sum_\alpha u_\alpha ^{\dagger }(\vec{r}%
)\, {\rm e}^{{\rm i}E_\alpha t}\, b_\alpha ^{\dagger
}+\displaystyle\sum_\alpha v_\alpha ^{\dagger }(\vec{r})\,{\rm
e}^{-{\rm i}E_\alpha t}\, d_\alpha \;,
\end{eqnarray}
where $u_\alpha (\vec{r})$ and $v_\alpha (\vec{r})$ form a complete
set of Dirac spinors in the coordinate space, and $b_\alpha $ and
$b_\alpha ^{\dagger }$ are the creation and annihilation operators
of a nucleon in the state $\alpha $. By $d_\alpha $ and $d_\alpha
^{\dagger }$ we denote the creation and the annihilation operators
for the anti-nucleons in the state $\alpha $. Similarly, the
$\sigma$ meson field may also be expanded in the following form:
\begin{equation}
\hat{\sigma}=\frac{1}{(2\pi)^{3/2}}\int
\displaystyle\frac{d^{3}k}{\sqrt{2 \omega_{\sigma}(k)}} \left[
c(\vec{k})\, {\rm e}^{{\rm i}\vec{k}\cdot \vec{r}}+
c^{\dagger}(\vec{k})\, {\rm e}^{-{\rm i}\vec{k}\cdot
\vec{r}}\right]\, .
\end{equation}
The omega field expansion, considering longitudinal and transverse
waves relative to the wave vector $\vec{k}$, reads:
\begin{eqnarray}
{\hat{\vec{\omega}}}=&\displaystyle\frac{1}{(2\pi)6^{3/2}}\int&\frac{d^{3}k}{\sqrt{
2\omega_{\omega}(k)}}\left\{\left[\frac{\omega_{\omega}}{m_{\omega}}
\, \vec{k}\,
a_{l}(\vec{k})+\sum_{t=1,2}\hat{e}_{t}(\vec{k})a_{t}(\vec{k})\right]{\rm
e}^{{\rm i}\vec{k}\cdot\vec{r}}+{\rm h.c} \right\}.
\end{eqnarray}
All creation and annihilation operators
$(c,c^{\dagger})$,$(a_{l},a_{l}^{\dagger})$ and
$(a_{t},a_{t}^{\dagger})$ obey canonical boson commutation
relations, and we have introduced the frequencies
$\omega_{\sigma}=\sqrt{m^{2}_{\sigma}+k^{2}}$ and
$\omega_{\omega}=\sqrt{m^{2}_{\omega}+k^{2}}$. Using the above
expansions, the free meson field Hamiltonians can be cast in the
form:
\begin{equation}
H_{\sigma}=\int d^{3}r\, \mathcal{H_{\sigma}}=\int d^{3}k\,
\omega_{\sigma}(k)c^{\dagger}(\vec{k})c(\vec{k}),\label{fsigma}
\end{equation}
and
\begin{equation}
H_{\omega}=\int d^{3}k\,
\omega_{\omega}(k)\left[a_{l}^{\dagger}(\vec{k})a_{l}(\vec{k})+
\sum_{t=1,2}a_{t}^{\dagger}(\vec{k})a_{t}(\vec{k})\right].\label{fomega}
\end{equation}
The nucleus state is assumed to be described by
$|\psi>=|A>|\sigma>|\omega>$, with $|A>$ representing an $A$ fermion
Slater determinant with the lowest energy states occupied (valence
or no-sea approximation), i.e.,
\begin{equation}
\mid A >=b_{\alpha _1}^{\dagger }b_{\alpha _2}^{\dagger }\cdots
b_{\alpha _A}^{\dagger }\mid 0>\;,\label{slater}
\end{equation}
where $\alpha _1,\dots ,\alpha _A$ are sets of single-particle
quantum numbers and $\mid 0>$ is the bare vacuum.  As it is usual in
$\sigma-\omega$ models, we work in the valence approximation, which
means that the polarization of the negative energy single particle
states is neglected. This is also a common approximation in
quark-meson chiral soliton models, such as the linear sigma model or
the chromodielectric model \cite{birsePPNP}. The approximation is
even more justifiable here since the binding energy per nucleon is
small compared with the rest mass of the nucleon. However, this
might not be the case in other approximations (see, for instance,
Ref. \cite{ring2}).

In the above product state, $|\sigma>$ represents a coherent state
describing the $\sigma$ mesons and $|\omega>$ a coherent state
describing the $\omega$ mesons. For instance, for the $\sigma$ meson
cloud:
\begin{eqnarray}
|\sigma>=N_{\sigma}\, {\rm exp} \left[\int d^{3}k\,
\eta(\vec{k})c^{\dagger}(\vec{k})\right]|0>\;,\label{sigma}
\end{eqnarray}
with $c(\vec{k})|\sigma>=\eta(\vec{k})|\sigma>$ and, from the
normalization of the state
\begin{eqnarray}
N_{\sigma}={\rm exp} \left[-\frac{1}{2}\int d^{3}k\,
|\eta(\vec{k})|^{2} \right].
\end{eqnarray}
We now enforce the mean value of the $\sigma$ field operator in the
coherent state to be equal to the potential obtained in the
mean-field Hartree approximation, i.e., we demand
\begin{eqnarray}
<\sigma|\hat{\sigma}|\sigma>=\phi_{0}(r),
\end{eqnarray}
for a spherical symmetric potential. This condition allows us to
determine, in an unique way, the function $\eta(\vec{k})$ in
(\ref{sigma}). Exploiting the spherical symmetry of the scalar
potential, one finds:
\begin{eqnarray}
\eta(k)=\sqrt{\frac{\omega_{\sigma}(k)}{\pi}}\int&dr\, r^{2}\,
j_{0}(kr)\phi_{0}(r)\;,\label{eta}
\end{eqnarray}
where $j_{0}$ is the spherical Bessel function of zeroth order. A
similar procedure can be carried out for the $\omega$ meson field.
In this case,
\begin{eqnarray}
|\omega>=&\displaystyle N_{\omega}\,\, {\rm exp}\left[ \int d^{3}k\,
[\Omega_{l}(\vec{k})a_{l}^{\dagger}(\vec{k})+\sum_{t=1,2}
\Omega_{t}(\vec{k})a_{t}^{\dagger}(\vec{k})\right]|0>\;,\label{omega}
\end{eqnarray}
and, using again the normalization and the properties of the vector
potential in the Hartree approximation, such as
$<\omega|\hat{\vec{\omega}}|\omega>=0$,
$<\omega|\hat{\omega}^{0}|\omega>=\omega_{0}(r)$ and
$<\omega|\hat{\vec{P}}_{\omega}|\omega>=\hat{r}\displaystyle\frac{d\omega_{0}(r)}{dr}$,
one finds:
\begin{eqnarray}
\Omega_{l}(k)=\frac{1}{m_{\omega}}\sqrt{\frac{\omega_{\omega}(k)}{\pi}}\int&dr
\, r^{2}j_{1}(kr)\displaystyle\frac{d\omega_{0}(r)}{dr},
\end{eqnarray}
and $\Omega_{t}(\vec{k})=0$. If we now take the states defined in
(\ref{slater}), (\ref{sigma}) and (\ref{omega}) and calculate the
mean value of the Hamiltonian obtained from (\ref{entot}), we
exactly recover the nucleus energy obtained in the usual Hartree
approximation.

Let us stress that the coherent state is a multiparticle
state and the description of meson clouds by coherent states
introduces many body correlations.

\section{The Center-of-Mass Approximate Projection}

Next, we want to obtain the center-of-mass (CM) correction to the
energy using the model described in section II. It is well known,
from the nuclear many-body theory, that mean-field approximations
break translational invariance (see Ref.~\cite{Greiner}) and that
the broken symmetry can be recovered  by applying the Peierls-Yoccoz
projection  to the symmetry-breaking state. The projection  operator
\begin{equation}
 {\cal P}_{\vec p}=\int {\rm exp}[{\rm i} (\hat{\vec{P}} -\vec{p})
 \cdot \vec{a}]\, d^3 \vec{a}\;,\label{py}
\end{equation}
exhibits the property
\begin{equation}
 {\cal P}_{\vec p} \, {\cal P}_{{\vec p} \prime}= \delta
 (\vec{p}-\vec{p}\, ' ){\cal P}_{\vec p}.
\end{equation}
In (\ref{py}),
$\hat{\vec{P}}=\hat{\vec{P_{A}}}+\hat{\vec{P_{\sigma}}}+\hat{\vec{P_{\omega}}}$
is the total linear momentum operator and $\vec{p}~$ the
corresponding eigenvalue. Our approach consists in assuming that
the model state representing the physical nucleus is obtained by
projecting the product mean-field Hartree state onto a zero
momentum ($\vec{p}=\vec{0}$) state (the procedure is known as
projection after variation). Since the Hamiltonian
$H=\int~d^{3}r\mathcal{H}$, with $\mathcal{H}$ given by
(\ref{entot}), commutes with the projection operator, we may write
the total energy, already corrected for the CM spurious motion, as
\begin{equation}
E_{\vec{p}=0}=\frac{<\psi \mid H{\cal P}_{\vec{p}={\vec 0}}\mid \psi >}{<\psi \mid {\cal P}_{\vec{%
p}={\vec 0}}\mid \psi >}\;.\label{etot}
\end{equation}
We emphasize that, in the valence approximation, the projection
operator acts on the mesons and on the positive energy fermions. The
vacuum single-particle states are unperturbed and the vacuum is
invariant under translations, so that the shifted states have the
same energy as the unshifted ones in the so-called variation before
projection method \cite{ringbook} that we are using here.

In order to compute the projected energy let us first consider
the norm overlap:
\begin{equation}
<\psi \mid {\cal P}_{\vec{p}={\vec 0}}\mid \psi
>=\int d \vec{a} <\sigma|{\rm e}^{{\rm i}\hat{\vec{P}}_{\sigma}\cdot\vec{a}}|\sigma>
<\omega|{\rm e}^{{\rm i}\hat{\vec{P}}_{\omega}\cdot\vec{a}}|\omega>
<A|{\rm e}^{i\hat{\vec{P}}_{A}\cdot\vec{a}}|A>\, ,
\end{equation}
and begin with the $\sigma$ field contribution. Its norm overlap
reads:
\begin{equation}
N_{\sigma}(a)=<\sigma|{\rm e}^{{\rm
i}\vec{P}_{\sigma}\cdot\vec{a}}|\sigma>={\rm
exp}\left\{4\pi\int~dk\, k^{2}|\eta(k)|^{2}[j_{0}(ka)-1]\right\},
\end{equation}
where $\eta(k)$ is defined by equation(\ref{eta}). We then find:
\begin{equation}
N_{\sigma}(a)={\rm exp} \left\{4\pi\int~dk\, k^{2}\,
\frac{(m_{\sigma}^{2}+k^{2})^{1/2}}{2}\tilde{\phi}_{0}^{2}(k)[j_{0}(ka)-1]\right\},
\end{equation}
where
\begin{equation}
\tilde{\phi}_{0}(k)=\int~dr\, r^{2}\, j_{0}(kr)\phi_{0}(r).
\end{equation}
Similarly, for the $\omega$ meson norm overlap
\begin{equation}
N_{\omega}(a)={\rm exp} \left\{\frac{4\pi}{m_{\omega}}\int~dk\,
k^{2}\,
\frac{(m_{\omega}^{2}+k^{2})^{1/2}}{2}\tilde{\omega}_{0}^{2}(k)\left[
j_{0}(ka)-1\right]\right\},
\end{equation}
where
\begin{equation}
\tilde{\omega}_{0}(k)=\int~dr\, r^{2}\, j_{0}(kr)\, \omega_{0}(r).
\end{equation}
The calculation of the fermionic part of the norm overlap is more
involved, and we just quote here the main result in a compact form:
\begin{equation}
N_{A}(a)=<A|{\rm e} ^{{\rm i} \hat{\vec{P}}_{A}\cdot\vec{a}}|A>=
{\rm det} \, B \,,
\end{equation}
where the $B$ matrix is defined by
\begin{equation}
B_{\alpha\beta}=<\alpha|\beta(a)>.
\end{equation}

Each label ($\alpha$ and $\beta$ stands for the set of particle
quantum numbers ($n,l,j,m$) as well as for the isospin
  projection quantum number necessary to classify the state.
  The ket $|\beta(a)>$ means a single-particle (four-component) state for which the spatial
  coordinate $\vec{r}$ is changed to $\vec{r}+\vec{a}$.

 Next, we move our attention to the energy kernel calculation. The
total Hamiltonian is written in the form:
\begin{equation}
H=H_{N}+H_{\sigma}+H_{\omega},\label{tot}
\end{equation}
where the first term contains the free fermion part as well as their
interaction with the $\sigma-\omega$ mesons. The second and third
terms are given by equations (\ref{fsigma}) and (\ref{fomega}) and
represent the free mesonic terms. Let us consider the free $\sigma$
field energy kernel. Using equations (\ref{sigma}),(\ref{eta}) and
the result:
\begin{equation}
|\sigma(a)>={\rm e} ^{{\rm i}
\hat{\vec{P}}_{\sigma}\cdot\vec{a}}|\sigma>=N_{\sigma}{\rm exp}
\left[~\int~d\vec{k}\, \eta^{'}(\vec{k})b(\vec{k})~\right]|0>,
\end{equation}
with $\eta^{'}(\vec{k})=\eta(\vec{k})\, {\rm e} ^{{\rm i}
\vec{k}\cdot\vec{a}}$, we obtain:
\begin{equation}
\varepsilon_{\sigma}(a)=<\sigma|H_{\sigma}|\sigma(a)>=\frac{1}{2}\left[\int~dk\,
k^{2}(m_{\sigma}^{2}+k^{2})\tilde{\phi}^{2}_0(k)j_{0}(ka)\right]N_{\sigma}(a)
\;.\label{enesigma}
\end{equation}
For the free $\omega$ meson energy kernel, a similar analysis leads
us to the following result:
\begin{equation}
\varepsilon_{\omega}(a)=<\omega|H_{\omega}|\omega(a)>=\frac{1}{2m_{\omega}^{2}}\left[\int~dk\,
k^{4}(m_{\omega}^{2}+k^{2})\tilde{\omega}_{0}^{2}(k)j_{0}(ka)
\right]N_{\omega}(a)\;.\label{eneomega}
\end{equation}
For the fermionic part of the energy kernel, it is more convenient
to rewrite the corresponding original Hamiltonian. From equation
(\ref{toth}), in the Hartree mean-field, we can read off the
fermionic Hamiltonian written in second quantized form:
\begin{equation}
H_{N}=\hat{h}^{(1)}+\hat{h}^{(1~2)}=\sum_{\alpha,\beta}h^{(1)}_{\alpha\beta}b^{\dagger}_{\alpha}b_{\beta}+
\sum_{\alpha,\beta\gamma\delta}h^{(1~2)}_{\alpha\beta\gamma\delta}
:b^{\dagger}_{\alpha}b_{\gamma} b^{\dagger}_{\beta}b_{\delta}:~,\label{fermionic}
\end{equation}
 with
\begin{equation}
h^{(1)}_{\alpha\beta}=\int~d\vec{r}\, u^{\dagger}_{\alpha}(\vec{r})
\left\{-{\rm i} \vec{\alpha}\cdot
\vec{\nabla}+\beta\left[M-g_\sigma\sigma(x)\right]+
\frac{g_\omega}{m_{\omega}^{2}}\vec{\nabla} \cdot
\vec{P_{\omega}}\right\}u_{\beta}(\vec{r})~,
\end{equation}
\noindent and
\begin{equation}
h^{(1~2)}_{\alpha\beta\gamma\delta}=\int\int~d\vec{r}d\vec{r}~'u^{\dagger}_{\alpha}(\vec{r})
u^{\dagger}_{\beta}(\vec{r}~')\frac{g_{\omega}^{2}}{m_{\omega}^{2}}\delta(\vec{r}-\vec{r}~')
u_{\gamma}(\vec{r})u_{\delta}(\vec{r}~').
\end{equation}

In the above equations, $u_{\alpha,\beta}$ represents the Dirac
single-particle spinor,  which we choose to be the Hartree
mean-field solution. Observing now that, the $\omega_0$ field should obey the Klein-Gordon
equation:
\begin{equation}
\nabla^{2}\omega_{0}(r)=-g_{\omega}\rho_{B}(r)+m_{\omega}^{2}\omega_{0}(r),
\end{equation}
\noindent and that $\vec{\nabla}\cdot
\vec{P}_{\omega}=-\nabla^{2}\omega_{0}(r)$, we may rewrite the
one-body part in (\ref{fermionic}) as:
\begin{equation}
\hat{h}^{(1)}=h_{\rm
MFA}-\frac{g_{\omega}^{2}}{m_{\omega}^{2}}\rho_{B}(\vec{r}),
\end{equation}
\noindent with:
\begin{equation}
h_{\rm MFA}u_{\alpha}=\epsilon_{\alpha}u_{\alpha}.
\end{equation}

We are now in position to perform the calculation of the fermionic
part of the energy kernel, which reads \cite{Alberto}:
\begin{equation}
\varepsilon_{N}(a)=<A|H_{N}{\rm e} ^{{\rm i}
\vec{P}_{A}\cdot\vec{a}}|A>=\sum_{\alpha}\epsilon_{\alpha}~N_{A}-<A|V^{(1)}{\rm
e} ^{{\rm i} \vec{P}_{A}\cdot \vec{a}}|A>+ <A|h^{(1~2)}{\rm e}
^{{\rm i} \vec{P}_{A}\cdot \vec{a}}|A>\, ,\label{nuckernel}
\end{equation}
\noindent where we have defined
$V^{(1)}=\frac{g_{\omega}^{2}}{m_{\omega}^{2}}\rho_{B}(\vec{r})$.
The second and third terms in equation (\ref{nuckernel}) can then be
obtained with the help of the well-known results (see, e.g., Ref.
\cite{brink}):
\begin{equation}
<A \mid V^{(1)}{\rm e} ^{{\rm i} \vec{P}_{A}\cdot\vec{a}}\mid A
>=N_{A}(a)\sum_{\alpha\beta}<\alpha\mid
V^{(1)}\mid\beta(a)>B^{-1}_{\beta\alpha}~,\label{1body}
\end{equation}
and
\begin{equation}
<A \mid h^{(1~2)}{\rm e} ^{{\rm i} \vec{P}_{A}\cdot\vec{a}}\mid
A>=\frac{1}{2}~N_{A}(a)\sum_{\alpha\beta\gamma\delta}<\alpha\beta\mid
h^{(1~2)}\mid\gamma(a)\delta(a)>B^{-1}_{\gamma\alpha}B^{-1}_{\delta\beta}~,\label{2body}
\end{equation}
\noindent where the exchange term has been neglected. Putting
everything  together, we finally obtain the total nucleus energy
corrected for the spurious CM motion
\begin{equation}
E_{\vec{p}=0}=\frac{\int~d\vec{a}~
[\varepsilon_{N}(a)N_{\sigma}(a)N_{\omega}(a)+
\varepsilon_{\sigma}(a)N_{A}(a)N_{\omega}(a)+\varepsilon_{\omega}(a)N_{A}(a)N_{\sigma}(a)]}
{\int~d\vec{a}\, [< \psi \mid \psi(a)>]}.\label{enetot}
\end{equation}
We stress that both the nucleons and the mesons were taken into
account in the evaluation of this projected energy.

\section{The Nuclear Charge Root-Mean-Square Radius}

We now turn to the evaluation of the nuclear root-mean-square (RMS)
radius in the formalism. Most of the measurements refer to the
proton charge RMS radius so we restrict ourselves to that case
(measurements for the neutron RMS radius are under way and are
receiving an increasing interest (\cite{parity})). On the other
hand, since the mesons in the model are all neutral we have to
consider just the nucleon (proton) contribution. Finally, in the
discussion below, we consider point-particle nucleons, though
nucleon form factors can be included without major difficulties.

The (translationally invariant) nuclear radius operator is

\begin{equation}
R^{2}_{\rm TI}=\sum_{i=1}^{A}e_{i}(\vec{r}_{i}-\vec{R}_{\rm
CM})^{2}~,
\end{equation}
where $\vec{R}_{\rm CM}$ is the center of mass coordinate and
$e_{i}$ is the charge of the i-th particle. The above operator can
be rewritten, for $N=Z$, as:
\begin{equation}
R^{2}_{\rm TI}=\frac{(A-1)}{A}\sum_{i=1}^{A}e_{i}\, r_{i}^{ 2}-
\frac{2}{A}\sum_{i<j}^{A}e_{i}\,
\vec{r}_{i}\cdot\vec{r}_{j}~.\label{rmsop}
\end{equation}

As for the energy calculation, the above radius operator commutes
with the total linear momentum, but our model wave function $|\psi>$
is not a total momentum eigenfunction, so the mean radius is then
given by:
\begin{equation}
<r^{2}>_{proj}=\frac{1}{Ze}\frac{\int~d\vec{a}\, <\psi|R^{2}_{\rm
TI}{\rm exp} ^{{\rm i} \vec{P}_{A}\cdot \vec{a}}|\psi>}
{\int~d\vec{a}\, < \psi \mid \psi(a)>}~.
\end{equation}

 Noting that the radius operator contains an one-body and a
two-body term, the numerator of the above equation can be worked
out with the help of equations like (\ref{1body}) and
(\ref{2body}) respectively.

\section{Numerical Applications for $N=Z$ Closed Shell Nuclei}

In order to perform applications to specific nuclei, we must solve
first the $\sigma-\omega$ model above described in the Hartree
approximation, disregarding the CM motion effects. This is totally
equivalent to solve the model treating the mesons as classical
fields \cite{Serot}. We choose to follow the method described in
reference \cite{ring}, where both the nucleon Dirac spinors and the
fields are expanded in three-dimensional harmonic oscillator
functions, $R_{kl}(r)$, and treat the expansion coefficients as
variational parameters. As we are dealing here with closed shell
nuclei only, we have:
\begin{equation}
g_{nlj}(r)=\sum_{k=0}^{N} C_k^{(nlj)} R_{kl}(r),
\end{equation}
\begin{equation}
f_{nlj}(r)=\sum_{k=0}^{N^{'}} \tilde{C}_k^{(nlj)} R_{kl}(r),
\end{equation}
with $g$ and $f$ being the up and lower radial components for the
single-particle wave function. For the meson fields:
\begin{equation}
B(r)=\sum_{k=0}^{N_{B}} C^{B}_k R_{k0}(r),
\end{equation}
where $B$ stands for $\phi_{0}$ or $\omega_{0}$ and $R_{kl}$ for the
radial harmonic oscillator function. Those expansions can be
introduced in the Dirac and Klein-Gordon equations and solved
self-consistently for the expansions coefficients $C_k$,
$\tilde{C}_k$ and $C^{B}_k$. After that, it is straightforward to
implement the calculation of the energy and RMS radius as presented
in the above sections, including the CM motion correction due to the
nucleons and mesons.

In Table I we show our results for the energy and for the
root-mean-square charge radius  without and with the CM projection
(the set of parameters for the nucleon and meson masses and for the
coupling constants are taken from reference \cite{HS}, but
disregarding the $\rho$ meson and the eletromagnetic field). In
Table II, we show the effect of the CM correction over the total
energy, without the meson contributions, i.e., only the nucleonic
degrees of freedom are taken in to account \cite{BJP}, together with
the usual harmonic oscillator approximation \cite{ring}, and also
including the correction computed from $<P_A^2/2AM>$ . From Table II
it is clear that the last  two corrections are similar  and not very
different from the Peierls-Yoccoz correction without the meson
degrees of freedom. Let us remember  that the Peierls-Yoccoz method
gives us not only the energy correction but also a translationally
invariant wave function for the system.

It is worthwhile to note that the inclusion of the mesonic
contribution makes the system too much bounded in comparison with
the case where just the fermionic contribution is explicitly taken
in to account. However, with a slight modification of  model
parameters, we are able to obtain reasonable results for the energy
and charge radius, as shown in Table III, in which the experimental
results are also displayed. For comparison within our calculation,
we have extracted the proton form factor contribution from the
experimental charge radius using the prescription given in equation
(6.2), ref. \cite{ring}.

 We must stress that the results shown in Table III are not
obtained from a careful fitting of the model parameters, which
should be done only after the inclusion of other mesons, as well
as non-linear terms in the original Lagrangian. Formally, these
terms can be readily  included but then the calculations become
more involved.

\section{Conclusions}
    We have computed the center-of-mass correction in the binding
energy and charge radius for spherical $N=Z$ nuclei using the well
known Peierls-Yoccoz projection method applied to the Hartree
solution of the Walecka $\sigma-\omega$ model. Although no
explicit reference has to be made to the mesonic states in the
Hartree approximation, we have chosen coherent states to describe
meson degrees of freedom. Those states are then completely
determined in this approximation and this allows us to  obtain the
nucleonic as well as the mesonic center-of-mass motion correction.
The numerical results show a very important contribution from the
mesons to the final binding energies and a modest but still
noticeable contribution to the charge radius, as compared to the
case where only the nucleonic CM correction is taken into account
or to the situation where no correction is done.  It is known that
the Peierls-Yoccoz projection suffers from the so-called mass
parameter problem which can be circumvented by using the
Peierls-Thouless or the so-called variation-after-projection
method \cite{ringbook}. Both are technically difficult to
implement but the latter might be feasible in systems of nucleons
and mesons, at least approximately. However, it was shown in Ref.
\cite{henley} that some observables , calculated in $\vec{p}=0$
states, do not suffer from the Peierls-Yoccoz mass problem. We
intend to perform a partial variation-after-projection in a
restricted meson space but do not expect large discrepancies for
oxygen and calcium, whose number of particles is already large, so
that quantum fluctuations are expected to be smaller.

    Another important feature of our result is the fact that the
CM correction, including the mesons, makes the system too much
bounded. This is expected as long as the model parameters were
chosen to reproduce some aspects of finite nuclei without that
correction. We have also shown that a few percent change in the
coupling constants can bring the total energy  and rms radius
close to the experimental values, at least for the $^{16}$O and
$^{40}$Ca cases. For $^{4}$He the results are still too far from
the desirable using our proposed values, but this is true even
when no CM correction is included and using the original
parametrization for that nucleus. Furthermore, the energy
correction in the $^{4}$He case is relatively large irrespective
to the approximation used to extract the CM motion, so we believe
that for this light mass region the projection after variation
procedure may not be applicable.

    In short, we may say that, if we want to take into account the
mesons in the center-of-mass correction applied to a relativistic
model for the nucleus, a new set of parameters must be found in
order to reproduce some basic nuclear properties as the binding
energy and radius. Once this is achieved, it would be interesting to
obtain other important nuclear properties as, e.g., the
eletromagnetic form factors and spectroscopic factors.  The model
and the techniques explored in this paper would provide a good
opportunity to obtain those observables. This work is in progress.

\vspace{0.30cm}

This work was partially supported by CNPq (Brazil) and GRICES
(Portugal).

\newpage
\begin{table}
\caption{Ground-state energy without ($E$) and with ($E_{\rm proj}$)
the CM correction for the three double-closed shell nuclei
considered in this work and the root-mean-square charge radius
without($<r^{2}>$) and with ($<r^{2}>^{1/2}_{\rm proj}$) the same
corrections.}
\begin{ruledtabular}
\begin{tabular}{ccccc}
$Nucleus$ & $E \, [ \,{\rm MeV}] $ & $E_{\rm proj} \, [\, {\rm
MeV}]$ & $<r^{2}>^{1/2} \, [ \, {\rm fm}]$
& $<r^{2}>^{1/2}_{\rm proj} \, [ \, {\rm fm}]$\\
\hline
$^{4}{\rm He}$  & $-4.85$   & $-68.95$ &  $2.06$ &  $1.84$ \\
$^{16}{\rm O}$  & $-94.63$  & $-190.67$ & $2.59$ &  $2.51$ \\
$^{40}{\rm Ca}$ & $-331.32$ & $-420.17$ & $3.33$ &  $3.28$ \\

\end{tabular}
\end{ruledtabular}
\end{table}

\newpage
\begin{table}
\caption{Ground-state energy, $E$, without the CM correction for
the three double-closed shell nuclei considered in this work and
with the CM correction $E_{\rm proj}$ but not considering the
meson degrees of freedom. Also shown is the energy with the CM
correction, $ E_{\rm harm}$, calculated in the harmonic oscillator
approximation and the energy corrected just by the subtraction of
$<P_A^2/2AM>$.}

\begin{ruledtabular}
\begin{tabular}{ccccc}
Nucleus & $E \, [ \,{\rm MeV}] $ & $E_{\rm proj} \, [\, {\rm MeV}]$
& $E_{\rm harm} \, [ \, {\rm MeV}]$ & $E_{P_A^2/2AM} [\rm{MeV}]$ \\
\hline
$^{4}{\rm He}$  & $  -4.85$  & $ -18.07$ & $ -24.22$ & $ -16.35$ \\
$^{16}{\rm O}$  & $ -94.63$  & $-107.87$ & $-106.83$ & $-104.92$\\
$^{40}{\rm Ca}$ & $-331.32$  & $-342.56$ & $-340.31$ & $-339.84$  \\
\end{tabular}
\end{ruledtabular}
\end{table}

\newpage
\begin{table}
\caption{Ground-state energy $E_{\rm proj}$ for the three
double-closed shell nuclei considered in this work  and charge
radius $<r^{2}>^{1/2}_{\rm proj}$ with the CM corrections
included, compared to the experimental results. The figures were
obtained using the values $g_s=10.45 $ , $g_v=13.82 $ and
$m_{s}=522$ MeV, as compared to the values $g_s=10.47 $,
$g_v=13.80$ and $m_{s}=522$ MeV from \cite{HS}.}

\begin{ruledtabular}
\begin{tabular}{ccccc}
$Nucleus$ & $ E_{\rm proj}\, [ \,{\rm MeV}] $ & $E_{\rm exp} \,
[\, {\rm MeV}]$ & $<r^{2}>^{1/2}_{\rm proj} \, [ \, {\rm fm}]$  &
$<r^{2}>^{1/2}_{\rm exp} \, [ \, {\rm fm}]$\\
\hline
$^{4}{\rm He}$  & $ -53.50$  & $ -28.30$ & $2.01$ & $1.57$ \\
$^{16}{\rm O}$  & $-158.50$  & $-127.68$ & $2.60$ & $2.61$ \\
$^{40}{\rm Ca}$ & $-339.14$  & $-338.00$ & $3.36$ & $3.39$ \\

\end{tabular}
\end{ruledtabular}
\end{table}

\end{document}